\documentclass[twocolumn,showpacs,preprintnumbers,pra,amsmath,amssymb]{revtex4}

\usepackage{graphicx}
\usepackage{bm}
\begin{document}
\title{How to Device-independently Generate States for Which Unambiguous State Discrimination is Impossible}

\author{Won-Young Hwang \footnote{Email: wyhwang@jnu.ac.kr}}

\affiliation{$^{1}$Department of Physics Education, Chonnam National University, Gwangju 500-757, Republic of Korea}

\begin{abstract}
We describe how to device-independently generate a set of quantum states for which unambiguous state discrimination is not possible. First, we derive a formula that a certain non-signaling black box must satisfy. Then, we describe how to generate a set of quantum states. Devices for generating states and possible measurements on the states can be put into a black box. Because this black box is non-signaling, it satisfies the formula. Using the formula, we prove that unambiguous state discrimination is not possible for the generated states. Because we did not consider internal mechanisms but only outcomes, our argument is valid for any (imperfect) devices.
\end{abstract}


\maketitle
\section{Introduction}
One of the most fundamental properties of quantum states that differentiates them from classical ones is that nonorthogonal states cannot be discriminated with certainty \cite{Ber04, Nie00}. Unambiguous state discrimination (USD) \cite{Iva87, Die88, Per88} has an all-versus-nothing nature. If the measurement is successful, states can be discriminated with certainty. However, the probability that the measurement fails is non-zero, which is a manifestation of the lack of certainty.

An intercept-resend attack using USD is an obstacle to quantum key distribution (QKD) protocols, especially to the Scarani-Ac\'in-Ribordy-Gisin 2004 protocol \cite{Sca04}. Thus, it is meaningful to avoid the USD attack. USD is applicable only for linearly independent states \cite{Che98}. Hence, a way to avoid the USD attack is to make it certain that the dimension of states generated is smaller than the number of states adopted in the protocol. The problem in practice is that unavoidable imperfection always exist in the source. Actual states generated deviate from the supposed ideal one. As a result, the dimension of the state deviates from that of the desired one, which makes the state vulnerable to USD attack. Here, if combined with high loss, even a very small imperfection can be dangerous for QKD users. Thus, we need a way to avoid the USD attack even with imperfections.

The purpose of this paper is to provide a method, with any source and measurement devices, to generate states for which the USD is impossible.
Because the method works with any source and measurement device, clearly it also works with an imperfect source and measurement device. What is guaranteed, however, is only the impossibility of USD. Nothing else, e.g., the identity of generated states, is guaranteed.

All device-independent quantum information processings so far, e.g., device-independent testing of device \cite{May04}, no-signaling QKD \cite{Bar05, Aci06}, and
device-independent QKD \cite{Aci07}, have been based on Bell's inequality violation \cite{Bel64}. However, our method is not based on Bell's inequality, but on the no-signaling principle. (Communication scenario similar to the one considered here was proposed by Herbert \cite{Her82} and was then applied to derive several bounds in quantum information, including approximate quantum cloning \cite{Gis98}, quantum state discrimination \cite{Hwa05, Bae11}, and quantum state estimation \cite{Han10}.)

\section{A no-signaling black box}
Assume that Alice and Eve, two persons, share a `black box' with the following properties; The part of the black box accessible to Alice has a knob that can be in two different directions. If a binary value $i$ Alice has chosen is $0$ ($1$), she turns the knob in one (the other) direction. Then, if she presses a button, the black box gives a binary outcome $j$ and an outcome $k$ ($k=0,1,2,...$), respectively, for Alice and Eve.
Here, we neither know nor care what happen inside the black box.
What we consider are only the outcomes. Let $P_i(j,k)$ denote joint probability, for knob choice $i$, that the outcomes are $j$ and $k$. Then, we get \cite{Cov91}
\begin{eqnarray}
P_i(\triangle,k)
&=& P_i(k|0) P_i(0,\triangle) + P_i(k|1) P_i(1,\triangle).
\label{1}
\end{eqnarray}
Here, $P_i(j,\triangle)$ and $P_i(\triangle,k)$ denote the marginal probabilities for Alice and Eve, respectively. (For clarity, we adopt a notation different from the one widely used in which the marginal distributions for Alice and Eve are $P_i(j)$ and $P_i(k)$, respectively.) $P_i(k|j)$ denotes the probability that Eve gets $k$ if Alice's outcome is $j$ for the knob choice $i$.
Note that the marginal probability $P_i(j,\triangle)$ can directly be estimated by using the outcomes of the black box for Alice.

The no-signaling principle implies that Eve's marginal distribution is independent of the knob choice:
\begin{eqnarray}
P_0(\triangle,k)= P_1(\triangle,k).
\label{2}
\end{eqnarray}
From Eqs. (\ref{1}) and (\ref{2}), we get
\begin{eqnarray}
  && P_0(k|0) P_0(0,\triangle)+ P_0(k|1) P_0(1,\triangle) \nonumber\\
&=& P_1(k|0) P_1(0,\triangle) + P_1(k|1) P_1(1,\triangle).
\label{3}
\end{eqnarray}
The condition in Eq. (\ref{3}) must be satisfied by any non-signaling black box.
\section{How to generate states}
Assume that each time Alice needs to generate a state chosen from among four states $|0\rangle, |1\rangle, |+\rangle, |-\rangle$ with equal probability. Here, $|0\rangle$ and $|1\rangle$ are two mutually orthogonal states and $|\pm\rangle = (1/\sqrt{2}) (|0\rangle \pm |1\rangle)$.

As is well known, if Alice can generate an ideal Bell state and can perform ideal measurements, she can generate ideal state:
Assume Alice has a Bell state
\begin{equation}
|\psi\rangle= \frac{1}{\sqrt{2}} (|0\rangle_{A_1} |0\rangle_{A_2} + |1\rangle_{A_1} |1\rangle_{A_2}),
\label{4}
\end{equation}
where $A_1$ and $A_2$ denote two separated parts in Alice's laboratory. Assume that $M_0$ ($M_1$) is a measurement composed of $\{|0\rangle \langle 0|, |1\rangle \langle 1|\}$ ($\{|+\rangle \langle +|, |-\rangle \langle -|\}$).
Alice chooses a value $i$ ($i=0,1$) with equal probability. She performs $M_i$ at location $A_1$. Then, at location $A_2$, the four states are generated with equal probability.

However, because of unavoidable imperfections, the actually generated state is slightly different from the ideal Bell state. The imperfect state is  denoted by $|\tilde{\psi}\rangle$. We also need to assume that the measurements are also imperfect. Let us denote the imperfect measurements by $\tilde{M_i}$.

Let us describe how to generate states for which USD is impossible device-independently. With imperfect devices, Alice does what she did in the ideal case. That is, she prepares a state $|\tilde{\psi}\rangle$. She randomly chooses a binary bit $i$. She performs a measurement $\tilde{M_i}$ at location $A_1$. If she performs $\tilde{M_0}$, she gets an outcome between $0$ and $1$ with a certain corresponding probability $p_1$ and $1-p_1$. When she gets $0$ ($1$), she regards a state at location $A_2$, whatever it actually is, as a $|0\rangle$ ($|1\rangle$) state. The imperfect states are denoted by $|\tilde{0}\rangle$ and $|\tilde{1}\rangle$, correspondingly.
Similarly, if she performs $\tilde{M_1}$, she gets an outcome between $0$ and $1$ with a certain corresponding probability $p_2$ and $1-p_2$. When she gets $0$ ($1$), she regards a state at location $A_2$, whatever it actually is, as a $|+\rangle$ ($|-\rangle$) state. The imperfect states are denoted by $|\tilde{+}\rangle$ and $|\tilde{-}\rangle$, correspondingly.
This way of generating the four states using an entangled state to hide basis information had been considered in Ref. 19 in the context of QKD without public announcement of basis \cite{Hwa98} and in Ref. 21 in the context of the effectiveness of the security proof.
\section{USD is not possible for the generated states}
Now let us argue why the set of the four imperfect states is not vulnerable to USD in any case. Let us consider a measurement $M$ by Eve with outcomes $k$ ($k=0,1,2,...$) performed on the imperfect state. Let us put all components, the state $|\tilde{\psi}\rangle$, Alice's measurement $\tilde{M_i}$, and Eve's one $M$, into a black box. By using a knob on the black box, Alice can choose between $\tilde{M_0}$ and $\tilde{M_1}$. Outcomes of the measurements are given by the black box. Now, we can see that the black box here is just the one we considered in Section II. We can see that the marginal probabilities for Alice, $P_i(0,\triangle)$ and $P_i(1,\triangle)$, are equal to $p_i$ and $1-p_i$, correspondingly.
Let $(0,0), (0,1), (1,0),$ and $(1,1)$ denote the imperfect states $|\tilde{0}\rangle$, $|\tilde{1}\rangle$, $|\tilde{+}\rangle$, and $|\tilde{-}\rangle$, respectively.
$P[k|(l,m)]$ is the conditional probability that outcome of $M$ is $k$ if the imperfect state is $(l,m)$. We can see that $P_l(k|m)$ is given by $P[k|(l,m)]$.
Since the black box here is non-signaling, it satisfies Eq. (\ref{3}), which is now
\begin{eqnarray}
p_0 P[k|(0,0)] &+& (1-p_0) P[k|(0,1)]= \nonumber\\
 && p_1 P[k|(1,0)]+ (1-p_1) P[k|(1,1)].
\label{5}
\end{eqnarray}

Now let us describe why Eq. (\ref{5}) implies that USD is not possible for the four states. In order that USD be possible, all conditional probabilities except for one should be zero. For example, $P[k|(0,0)]=1$ and $P[k|(0,1)]= P[k|(1,0)]= P[k|(1,1)]=0$ means that if the outcome is $k$, the measured state must be $(0,0)$ by Bayes's formula.
Clearly, we have
\begin{eqnarray}
&&\min(p_0, 1-p_0)\{P[k|(0,0)]+ P[k|(0,1)]\} \nonumber\\
&&\leq p_0 P[k|(0,0)]+ (1-p_0) P[k|(0,1)],
\label{6}
\end{eqnarray}
where $\min(p_0, 1-p_0)$ is the minimal value between $p_0$ and $1-p_0$. Similarly, we obtain
\begin{eqnarray}
&& p_1 P[k|(1,0)]+ (1-p_1) P[k|(1,1)] \nonumber\\
&&\leq \max(p_1, 1-p_1)\{P[k|(1,0)]+ P[k|(1,1)]\},
\label{7}
\end{eqnarray}
where $\max(p_1, 1-p_1)$ is the maximal value between $p_1$ and $1-p_1$. Combining Eqs. (\ref{5})-(\ref{7}), we get
\begin{eqnarray}
&&P[k|(1,0)]+ P[k|(1,1)] \geq \nonumber\\
&&\frac{\min(p_0, 1-p_0)}{\max(p_1, 1-p_1)}\{P[k|(0,0)]+ P[k|(0,1)]\}.
\label{8}
\end{eqnarray}
Similarly, we also obtain
\begin{eqnarray}
&&P[k|(0,0)]+ P[k|(0,1)]
\geq \nonumber\\
&&\frac{\min(p_1, 1-p_1)}{\max(p_0, 1-p_0)} \{P[k|(1,0)]+ P[k|(1,1)]\}.
\label{9}
\end{eqnarray}
Using Eqs. (\ref{8}) and (\ref{9}), we can derive a bound on the Eve's probability to make a correct guess on the identity of states. Note that Alice's marginal distributions can be statistically determined by using the outcomes for Alice. Let us consider the case when $p_i= 1/2$, as an example. (The case is obtained when the ideal Bell state was employed. However, our argument does not depend on how it was obtained. Our argument is valid as long as $p_i$ is observed to be $1/2$.) In this case, Eqs. (\ref{8}) and (\ref{9}) give
\begin{eqnarray}
P[k|(0,0)]+ P[k|(0,1)]= P[k|(1,0)]+ P[k|(1,1)].
\label{10}
\end{eqnarray}
Here we can see USD is not possible: for any outcome $k$, Eve's probability to make a correct guess on the identity of states is bounded by $1/2$.
\section{Discussion and Conclusion}
Equations (\ref{8}) and (\ref{9}) imply that the more deviated from $1/2$ the $p_i$ is, the higher the bound on Eve's probability to make a correct guess is. Let us consider an extreme case when $p_0=0$. Now, $P[k|(0,0)]=1$ and $P[k|(0,1)]= P[k|(1,0)]= P[k|(1,1)]=0$ are not excluded. However, this does not allow USD because the state $(0,0)$ is not even generated.

Note that, by the no-signaling principle, Eq. (\ref{5}) is satisfied in any case. The state $|\tilde{\psi}\rangle$ and Alice's measurement and Eve's measurement need not even be close to the ideal ones. Our argument for the impossibility of USD is still valid for this case. However, the generated states are not proper for QKD protocols because they deviate greatly from the ideal ones.

In conclusion, we derived a formula that a certain non-signaling black box must satisfy. Then, we described how to generate a set of quantum states. Devices for generating states and possible measurements on the states can be put into a black box. Because this black box is non-signaling, it satisfies the formula. Using the formula, we proved that USD is not possible for the generated states. Because we did not consider internal mechanisms but only outcomes, our argument is valid for any (imperfect) devices.

\section*{Acknowledgment}
This study was supported by the Basic Science Research Program through the National Research Foundation of Korea (NRF) funded by the Ministry of Education, Science and Technology (2010-0007208).


\begin{references}
\bibitem{Ber04} J. A. Bergou, U. Herzog, and Mark Hillery, Lect. Notes Phys. {\bf 649}, 417 (2004).
\bibitem{Nie00} M. A. Nielsen and I. L. Chuang, {\it Quantum Computation and Quantum Information}, (Cambridge
                University Press, Cambridge, U.K., 2000.)
\bibitem{Iva87} I. D. Ivanovic, Phys. Lett. A {\bf 123}, 257 (1987).
\bibitem{Die88} D. Dieks, Phys. Lett. A {\bf 126}, 303 (1988).
\bibitem{Per88} A. Peres, Phys. Lett. A {\bf 128}, 19 (1988).

\bibitem{Sca04} V. Scarani, A. Ac\'in, G. Ribordy, and N. Gisin, Phys. Rev. Lett. {\bf 92}, 057901 (2004).
\bibitem{Che98} A. Chefles, Phys. Lett. A {\bf 239}, 339 (1998).
\bibitem{May04} D. Mayers and A. Yao, Quant. Inf. Comput. {\bf 4}, 273 (2004).
\bibitem{Bar05} J. Barrett, L. Hardy, and A. Kent, Phys. Rev. Lett. {\bf 95}, 010503 (2005).
\bibitem{Aci06} A. Ac\'in, N. Gisin, and L. Masanes, Phys. Rev. Lett. {\bf 97}, 010503 (2006).
\bibitem{Aci07} A. Ac\'in, N. Brunner, N. Gisin, S. Massar, S. Pironio, and V. Scarani, Phys. Rev. Lett. {\bf 98}, 230501 (2007).
\bibitem{Bel64} J. S. Bell, Physics {\bf1}, 195 (1964).
\bibitem{Her82} N. Herbert, Found. Phys. {\bf 12}, 1171 (1982).
\bibitem{Gis98} N. Gisin, Phys. Letts. A {\bf 242}, 1 (1998).
\bibitem{Hwa05} W-Y. Hwang, Phys. Rev. A {\bf 71}, 062315 (2005).
\bibitem{Bae11} J. Bae, W-Y. Hwang, and Y-D. Han, Phys. Rev. Lett. {\bf 107}, 170403 (2011).
\bibitem{Han10} Y-D. Han, J. Bae, X-B. Wang, and W-Y. Hwang, Phys. Rev. A {\bf 82}, 062318 (2010).
\bibitem{Cov91} T. M. Cover and J. A. Thomas, {\it Elements of Information Theory} (John Wiley and Sons, Inc., 1991).
\bibitem{Hwa03} W-Y. Hwang, X-B. Wang, K. Matsumoto, J. Kim, and H-W. Lee, Phys. Rev. A {\bf 67}, 012302 (2003).
\bibitem{Hwa98} W. Y. Hwang, I. G. Koh, and Y. D. Han, Phys. Lett. {\bf 244}, 489 (1998).
\bibitem{Koa03} M. Koashi and J. Preskill, Phys. Rev. Lett. {\bf 90}, 057902 (2003).
\end{references}
\end{document}